\documentclass[angeo]{copernicus}

\usepackage{amsmath}

\newcommand{\diffp}[2]{\frac{\partial #1}{\partial #2}}

\begin{document}

\title{Apparent temperature anisotropies due to wave activity in the solar wind}

\author[1,2]{D.~Verscharen}
\author[1]{E.~Marsch}

\affil[1]{Max Planck Institute for Solar System Research, Max-Planck-Str. 2, 37191 Katlenburg-Lindau, Germany}
\affil[2]{Institute for Theoretical Physics, TU Braunschweig, Mendelssohnstr. 3, 38106 Braunschweig, Germany}

\runningtitle{Apparent temperature anisotropies}

\runningauthor{D.~Verscharen and E.~Marsch}

\correspondence{Daniel Verscharen\\ (verscharen@mps.mpg.de)}

\received{16. March 2011}
\revised{}
\accepted{}
\published{}


\firstpage{1}

\maketitle

\begin{abstract}
The fast solar wind is a collisionless plasma permeated by
plasma waves on many different scales. A plasma wave represents the natural
interplay between the periodic changes of the electromagnetic field and the
associated coherent motions of the plasma particles. In this paper, a model
velocity distribution function is derived for a plasma in a single, coherent, large-amplitude wave. This model allows one to study the kinetic
effects of wave motions on particle distributions. They are by in-situ
spacecraft measured by counting, over a certain sampling time, the particles
coming from various directions and having different energies. We compare our
results with the measurements by the Helios spacecraft, and thus find that by
assuming high wave activity we are able to explain key observed features of
the measured distributions within the framework of our model. We also address
the recent discussions on nonresonant wave--particle interactions and apparent
heating. The applied time-averaging procedure leads to an apparent ion
temperature anisotropy which is connected but not identical to the intrinsic
temperature of the underlying distribution function.
\end{abstract}

\introduction

It has been known for a long time that the solar wind is a turbulent plasma
with wave activity occurring on a wide range of different scales
\citep[][]{belcher71,tu95,horbury05}. Of course, any plasma wave has
to fulfill the Vlasov-Maxwell equations, and it can therefore be understood
as the space- and time-dependent self-consistent interplay between the
periodic variations of the electromagnetic field and related motions of the
particles, being represented by their velocity distribution function (VDF).

In the recent literature, the shaping of distribution functions due to wave
activity has been widely discussed \citep{wang06,wu07,wu09,wang09}.
Obviously, the presence of wave forces (or their spectra) will lead to a
deformation of the distribution function with respect to a Maxwellian shape
and cause a velocity spreading after appropriate averaging over the wave
effects. The plasma-physics definition of temperature is based on a
statistical particle ensemble and usually defined in terms of the random
kinetic energy (via the mean square of the particle velocity) in the
particles' mean-velocity frame. This definition of temperature is not
necessarily equal to the thermodynamic temperature, which may be called the
\emph{intrinsic temperature} of the particle ensemble. The mean square
velocity fluctuation owing to wave activity is able to cause an effective
broadening of the distribution function similar to real heating, and thus may
mimic genuine heating. Therefore some authors have referred to this process
as \emph{apparent heating} \citep{wang06}, others as \emph{nonresonant
wave--particle interactions}. The common ground of these wave effects is that
they are reversible, and therefore not dissipative, and do not represent real
heating. Collisions, however, might be able to dissipate coherent wave motion
efficiently, and thus will lead to a real heating and an increase in the
intrinsic temperature \citep{schekochihin08,howes08}. In the following,
collisions are excluded from the treatment of the VDFs, in order to
demonstrate the collisionless effects of waves and reveal the apparent
heating due to wave activity.

Substantial ion temperature anisotropies have been observed in the solar wind
and discussed by different authors
\citep{marsch81,marsch04,bale09,bourouaine10}. Typically the proton
temperature is higher perpendicular to the magnetic field than parallel to
it. These anisotropies have mostly been discussed as being the result of the
cyclotron-resonant interaction with circularly polarized waves, a process
which can be quantified by means of quasi-linear theory
\citep[e.g.][]{akhiezer75,heuer07}. On the other hand, such anisotropic
distribution functions can become unstable if the anisotropy exceeds a certain threshold that depends on the plasma beta \citep{gary00,gary01}. For a typical solar wind beta of about 1, the beta-dependence is not severe and the distribution function becomes unstable for $T_{\perp}/T_{\parallel}\gtrsim 2$. This instability can in turn excite and radiate ion-cyclotron waves. In this way, wave excitation can reduce the ion
temperature anisotropy efficiently and yield moderate and stable values
\citep{gary93,bale09}.

The general theory of this wave--particle interaction is well established, and
in fact many traits of it were confirmed by observations \citep{marsch06}.
However, the apparent heating effects are not well understood and have not
yet been discussed in the context of ion--wave interactions below and near the
ion inertial scale. Yet, they should be included in an appropriate
description of space plasmas that are subjected to strong
Alfv\'en/ion-cyclotron wave activity, such as it is typical for the solar
wind in the inner heliosphere. To study possible physical causes of apparent
wave heating is the aim of the present paper. First, we discuss the effect of
a strong plasma wave on an intrinsically Maxwellian distribution function.
Second, a concise form of the resulting model VDF is constructed, and thus we
can discuss how such a distribution function would look like in a real plasma
measurement made on a space probe.

\section{Model distribution functions}

\subsection{Wave effects on the distribution function}

To study the effects that waves have on the shape of a velocity distribution
function, we have to determine the constants of individual particle motion in
a given wave field, and then exploit the fact that any function of these
constants of motion is a solution of the Vlasov equation
\citep{davidson83,stix92}. Here, we discuss the influence of a single
monochromatic wave only. It is supposed here to be transversal and left-hand
circularly polarized, and its magnetic field can be assumed to have the form
\begin{equation}\label{bfielddef}
\vec B=\begin{pmatrix}b\cos(kz-\omega t)\\ b\sin(kz-\omega t)
\\B_0\end{pmatrix},
\end{equation}
with the constant field component $B_0$ along the $z$-axis and a wave
amplitude $b$. The wave frequency is $\omega$ and the parallel wavenumber $k$,
and its magnetic field is associated with an electric field which according
to Faraday's law is given by
\begin{equation}
\nabla \times \vec E=-\frac{1}{c}\diffp{\vec B}{t}.
\end{equation}
The equation of motion for a single particle with charge $q$ and mass $m$ in
this electromagnetic field is determined by the Lorentz force:
\begin{equation}\label{eqmotion}
\frac{\mathrm d \vec v}{\mathrm d t}=\frac{q}{m}\left(\vec E+\frac{1}{c}\vec
v\times \vec B\right).
\end{equation}
It is useful and transparent to write this equation in the components of a
cylindrical coordinate system (symmetry around the $z$-axis) as follows:
\begin{align}
\frac{\mathrm d v_{\perp}}{\mathrm dt}&=
\Omega \left(\frac{\omega}{k}-v_{\parallel}\right)\frac{b}{B_0}\sin\left(\phi-\varphi\right) \label{dvdt1},\\
\frac{\mathrm d v_{\parallel}}{\mathrm dt}&=
\Omega v_{\perp}\frac{b}{B_0}\sin\left(\phi-\varphi\right), \\
\frac{\mathrm d \varphi}{\mathrm dt} &=
-\Omega\left[1+\left(\frac{\omega}{k}-v_{\parallel}\right)\frac{1}{v_{\perp}}\frac{b}{B_0}
\cos\left(\phi-\varphi\right)\right],\qquad v_{\perp}\neq 0 \label{dvdt3},
\end{align}
where we made use of the abbreviation $\phi=kz-\omega t$ for the wave phase.
The gyrofrequency is denoted by $\Omega=qB_0/(mc)$. A similar set of equations  has already been used in a test-particle description to describe the nonlinear behavior of particles that are trapped in the wave fields \citep {matsumoto79}. Also nonresonant heating effects have been treated with similar equations under the assumption of low plasma betas both in the monochromatic parallel case and for a spectrum of oblique MHD waves \citep{hamza06,lu07,li07,lu09}. However, the initial conditions in these cases break the condition of the coherent particle motion that is necessary to maintain the wave in a self-consistent way. Neglecting the coherence and violating the self-consistency may be an appropriate description for a minor particle species in the sense of a test-particle approach. But the description is insufficient for the dominating main species that carry the currents and charges maintaining the wave itself. \citet{li07} also consider consistent initial conditions that do reflect the coherent wave motion and find that these particles do not experience the non-resonant heating because they are not picked-up by the wave fields. A model for the coherent particle motion of the dominating species in a wave field at high plasma betas should be based on a Vlasov description to take the finite thermal width of the distribution into account. Numerical self-consistent simulations are another approach to this problem \citep[e.g.][]{li05,araneda08,araneda09,maneva10}. In order to determine an adequate model distribution function, we first can determine two constants of motion for the kinematic system from Eqs.~(\ref{dvdt1}-\ref{dvdt3}). These are the
generalized momentum of the particle and its total kinetic energy in the wave
frame:
\begin{align}
M&=v_{\perp}\cos\left(\phi-\varphi\right)+\frac{B_0}{b}v_{\parallel}
\left[1-\frac{\omega}{\Omega}+\frac{kv_{\parallel}}{2\Omega}\right],\\
P&=v_{\perp}^2+\left(v_{\parallel}-\frac{\omega}{k}\right)^2.\label{Pconstant}
\end{align}
Both $M$ and $P$ can be shown to be constant, simply by taking the
derivatives with respect to $t$ and using the equations of motion in
Eqs.~(\ref{dvdt1}-\ref{dvdt3}). It is known \citep{akhiezer751} that any
distribution function which is a function of these constants of motion always
fulfills the Vlasov equation. Therefore, we can make the ansatz
\begin{equation}
f=N\exp\left(-\frac{P}{v_{\mathrm{th}}^2}\right)\exp\left(\frac{2V_{\mathrm
w} M}{v_{\mathrm{th}}^2}\right),
\end{equation}
with the normalization factor $N$ and constant coefficients $v_{\mathrm{th}}$
and $V_{\mathrm w}$, which we can essentially define as the density, thermal
velocity of the VDF and mean fluid-velocity amplitude of the particles in
association with the wave motion.

The right second term in the expression for $M$ compensates the frame-shift
of $v_{\parallel}$ by $\omega/k$ in the definition of $P$, if the condition
\begin{equation}\label{polarizationvw}
V_{\mathrm w}=-\frac{b}{B_0}\frac{\omega/k}{1-\omega/\Omega}
\end{equation}
is fulfilled, and if we can ignore the weak effects due to a small spread in
the parallel direction ($kv_{\parallel}\ll \Omega$). Interestingly enough,
this relation then corresponds to the wave polarization relation found by
\citet{sonnerup67} for a circularly polarized wave with a vanishing parallel
bulk drift. In their classical solution, they showed that the transversal
velocity is determined by
\begin{equation}
\vec V_{\mathrm t}=-\frac{\omega/k}{1-\omega/\Omega}\frac{\vec B_{\mathrm t}}{B_0}
\end{equation}
for vanishing drifts in the $z$-direction. We define the transversal magnetic
field vector as $\vec B_{\mathrm t} = (B_x,B_y,0)$, with the first two
components as obtained from Eq.~(\ref{bfielddef}). The dispersion relation of
the waves was also given in that work and can be written as
\begin{equation}
k^2+\sum \limits_j \frac{1}{\ell_j^2}\frac{\omega}{\omega-\Omega_j}=0,
\end{equation}
whereby the small displacement current in Maxwell's equations was neglected.
The index $j$ numbers all participating species (in the case considered later
only protons and electrons), and $\ell_j=c/\omega_j$ is the corresponding ion
inertial length, with the species' plasma frequency $\omega_j=\sqrt{4\pi
n_jq_j^2/m_j}$.

We find that after normalization the non-gyrotropic model VDF of the
particles in response to the wave forces reads
\begin{multline}\label{fwave}
f_{\mathrm w}(v_{\perp},
v_{\parallel},\varphi)=\frac{n_0}{\pi^{3/2}v_{\mathrm{th}}^3 }
\exp\left(-\frac{V_{\mathrm w}^2+\omega^2/k^2}{v_{\mathrm{th}}^2}\right)\times \\
\times \exp\left(-\frac{v_{\perp}^2+v_{\parallel}^2}{v_{\mathrm{th}}^2}\right)
\exp\left(\frac{2v_{\perp}V_{\mathrm
w}}{v_{\mathrm{th}}^2}\cos\left(\phi-\varphi\right)\right).
\end{multline}
The first exponential stems from the normalization and does not change the
structure of the distribution function, but depends on the particles sloshing
velocity amplitude, $V_{\mathrm w}$, and wave phase speed, $\omega/k$. It is
interesting to note, that this model distribution function is equal to a
Maxwellian distribution in cylindrical coordinates, yet which is shifted by
$\vec V_{\mathrm t}$ in the transversal direction and can be written in this
shifted Maxwellian form as
\begin{equation}
f\sim\exp\left(-\frac{(v_x-V_x)^2+(v_y-V_y)^2}{v_{\mathrm{th}}^2}\right),
\end{equation}
with the cartesian speed components $V_x$ and $V_y$ reflecting the rigid
displacement of the whole VDF in the wave field. This is an ansatz commonly
used to initialize consistently numerical simulations \citep{araneda08}, and
represents the sloshing motion of particles in the wave field
\citep[e.g.][]{markovskii09}. 

A VDF that has an additional intrinsic temperature anisotropy is supposed to be represented by a modified bi-Maxwellian distribution.  The appropriate choice for a VDF in a wave field including an intrinsic temperature anisotropy is given by
\begin{multline}
f_{\mathrm a}\sim 
\exp\left(-\frac{V_{\mathrm w}^2}{v_{\mathrm{th}\perp}^2}-\frac{\omega^2/k^2}{v_{\mathrm{th}\parallel}^2}\right)
\exp\left(-\frac{v_{\perp}^2}{v_{\mathrm{th}\perp}^2}-\frac{v_{\parallel}^2}{v_{\mathrm{th}\parallel}^2}\right)\times \\
\times \exp\left(\frac{2v_{\perp}V_{\mathrm
w}}{v_{\mathrm{th}\perp}^2}\cos\left(\phi-\varphi\right)\right),
\end{multline}
where different thermal speeds in the perpendicular and parallel direction are chosen to account for the intrinsic anisotropy. It is important to note, however, that this distribution function is not an exact solution of the Vlasov equation anymore. Without wave activity, this distribution function obtains the usual bi-Maxwellian form. We do not apply this modified bi-Maxwellian VDF in the following since we focus our considerations on the role of apparent temperature anisotropies only.

\subsection{Wave effects on particle measurements}

The solar wind is permeated by magnetic field fluctuations and waves
\citep{tu95} of all kind. A particle detector that is able to determine
the velocity distribution function of particles (e.g., such as flown onboard
the Helios spacecraft) counts particles in different energy and direction
channels, and thereby integrates the net particle fluxes into the various
single channels over the so-called sampling time $T$. In a first
approximation, we may interpret this time as kind of an exposure time (like
in photography), and thereby neglect time-dependent sampling effects on the
instruments pointing directions (to different solid looking angles), or on
the accessibility of the particles to the different energy channels during
the measurement cycle. Effects of the proper motion of the detector with
respect to the solar wind flow are also neglected for the first estimation. A
possibility to handle the effects arising from this relative motion would be
to treat the waves as frozen in the solar wind and being with the fixed
spatial structure convected over the space probe. This assumption is called
Taylor hypothesis and is valid only for $\vec k \cdot \vec
V_{\mathrm{SW}}\gg \omega$, where $\vec V_{\mathrm{SW}}$ denotes the solar
wind flow velocity. Since the Alfv\'en speed is typically about a factor of 10 less than the flow speed of the solar wind in the spacecraft reference frame, this Doppler effect would even increase the sampling problem because waves with lower wavenumbers appear at higher frequencies for the detector. These lower wavenumber structures typically have a higher power than waves at higher wavenumbers. Therefore, the relative motion of the solar wind with respect to the spacecraft amplifies the wave effect on the measured distribution function additionally.

Such a model instrument would not be able to take snapshots of the VDF but
integrate the sloshing distribution over time $T$, and would therefore obtain
a spread in the VDF due to the wave activity. In the following, we determine
theoretically the influence of this final integration time on the actual
measurement.

The relevant spectral range is limited to frequencies that are higher than
the sampling frequency, $2\pi/T$, because slower motions would be more or
less resolved. They would merely lead to a rigid shift of the full
distribution function without deformation. The analysis software of a plasma
instrument would set the origin of the reference frame to the shifted center
of the VDF, so that no change would be detectable. On the upper frequency
side, the acting part of the wave spectrum should be limited by the
gyrofrequency, because the waves are supposed to be strongly damped in the
case of Alfv\'en-cyclotron waves at this scale, and thus beyond it the
observed spectral energy goes down significantly. The slope of the spectral
energy density follows Kolmogorov's law at lower frequencies
\citep{tu95}. Numerical simulations show that at wave numbers around
$k\approx 0.8/\ell_{\mathrm p}$ the Alfv\'en/ion-cyclotron spectral slope
usually breaks because of the onset of dissipation at these scales
\citep{ofman05}. Here, we only assume a monochromatic wave
with $\omega=2\pi/T$. 

Given all these assumptions, the time-averaging process of the VDF under the
influence of waves, $f_{\mathrm w}$, may be expressed mathematically as
\begin{equation}\label{fbarbegin}
\bar f=\frac{1}{T}\int \limits_{0}^{T} f_{\mathrm w}\mathrm dt.
\end{equation}
We can now insert our model function of Eq.~(\ref{fwave}), in which the last
exponential factor can be expanded \citep{abramowitz72} and represented by a
sum over modified Bessel functions according to the relation:
\begin{equation}
e^{a\cos b}=I_0(a)+2\sum\limits_{m=1}^{\infty}I_m(a)\cos(mb).
\end{equation}
The time dependence of the averaged distribution function $\bar f$ is hidden
in the wave phase, $\phi(t)$. The coordinates of the distribution function
(i.e., $z$, $v_{\perp}$, and $v_{\parallel}$) have no time dependence in this
context. The expression for $f_{\mathrm w}$ in Eq.~(\ref{fwave}) delivers a value of the
VDF at each position in phase space and at each time. Without any
restriction, the spatial position can be taken $z=0$. Then the time-averaged
distribution function can be written as
\begin{multline}\label{fbargen}
\bar
f=N\exp\left(-\frac{v_{\perp}^2+v_{\parallel}^2}{v_{\mathrm{th}}^2}\right)
\left[I_0\left(\frac{2v_{\perp}V_{\mathrm w}}{v_{\mathrm{th}}^2}\right)\right.+\\
\left.
+\frac{2}{T}\sum\limits_{m=1}^{\infty}I_m\left(\frac{2v_{\perp}V_{\mathrm
w}}{v_{\mathrm{th}}^2}\right)\frac{\sin (m\omega T+m\varphi)-\sin(m\varphi)}{
m\omega }\right]
\end{multline}
We find that the sum is always zero for all possible positions in $\varphi$
if $\omega=2\pi/T$. So we can simplify the above formula and write
\begin{equation}\label{fbar1}
\bar f_1=N\exp\left(-\frac{v_{\perp}^2+v_{\parallel}^2}{v_{\mathrm{th}}^2}
\right)I_0\left(\frac{2v_{\perp}V_{\mathrm w}}{v_{\mathrm{th}}^2}\right),
\end{equation}
and thus we see that the intrinsically non-gyrotropic distribution function
appears to become gyrotropic again after this kind of averaging process. A
wider spectrum of waves could lead to a deformation of the distribution
function still in the coordinate $\varphi$, but as stated above the higher
frequencies on shorter time scales than the sampling time $T$ are not
expected to change the result much.

\begin{figure}[t]
\vspace*{2mm}
\begin{center}
\includegraphics[width=8.3cm]{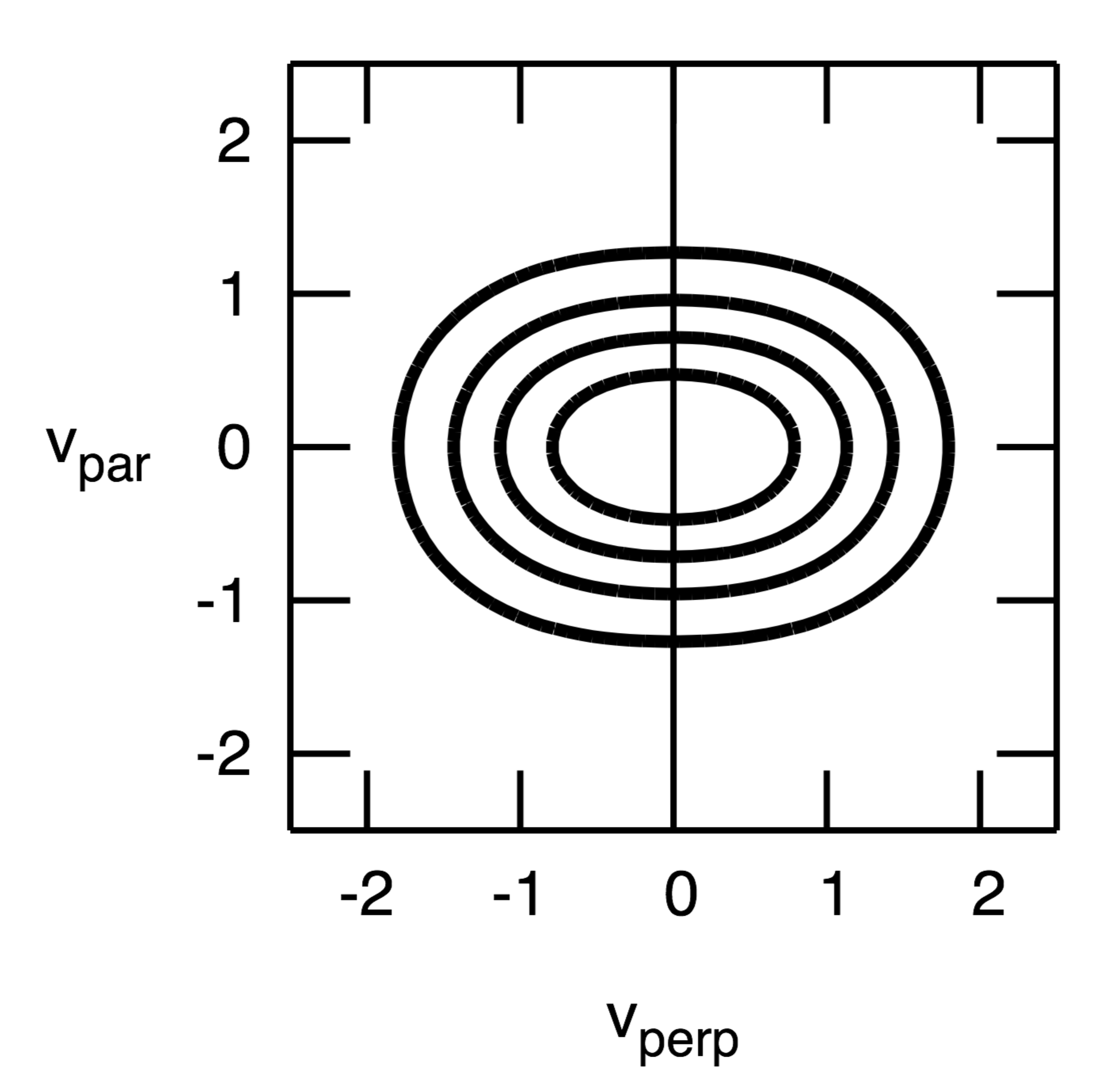}
\end{center}
\caption{Distribution function in the presence of a large-amplitude wave. The solid line
represents the magnetic field direction. The velocities are given in units of
the thermal speed. The broadening in the perpendicular direction is clearly
visible. } \label{fig_final_linear_deformed_sw}
\end{figure}

We assume next an Alfv\'en-cyclotron wave. This wave has to fulfill a
dispersion relation, which can be taken from the cold plasma limit, yielding
the dispersion relation as
\begin{equation}
\left(\frac{\omega}{\Omega_{\mathrm p}}\right)^2=(k\ell_{\mathrm
p})^2+\frac{1}{2}(k\ell_{\mathrm p})^4-\frac{1}{2}(k\ell_{\mathrm
p})^3\sqrt{(k\ell_{\mathrm p})^2+4}
\end{equation}
\citep{stix92,chandran10b}. For high values of $k$, the dispersion shows the asymptotical behavior $\omega\rightarrow \Omega_{\mathrm p}$, which can be seen by expanding the function $\sqrt{1+x}$ with $x=4/(k\ell_{\mathrm p})^2$ to second order. Typical solar wind parameters are used for a
distance between spacecraft and sun of about 0.5 AU, which is where strongly
non-Maxwellian distributions and high wave activity were usually observed in
fast solar wind by the Helios spacecraft. The plasma beta is set to 0.1 and
the sampling time to 10~s, which is the real sampling time of the Helios spacecraft. The constant background magnetic field $B_0$ is
assumed to be $5\times 10^{-4}\,\mathrm G$, and the proton density to be
$n=10\,\mathrm{cm}^{-3}$. The relative wave amplitude is set to $b/B_0=0.25$.
The distribution function at $\varphi=\pi$ is plotted at negative values for
$v_{\perp}$, in order to make plots that can more easily be compared with the
observations. The calculated distribution function is shown in
Fig.~\ref{fig_final_linear_deformed_sw}.

Heavy ions, which are also present in the solar wind \citep{vonsteiger08}, react on the wave field in a slightly different way. The dependence of $V_{\mathrm w}$ on the charge-to-mass ratio in Eq.~(\ref{polarizationvw}) is small for lower frequencies. The thermal speed, however, is significantly lower than the proton thermal speed by a factor of $\sqrt{m_{\mathrm p}/m_{\mathrm i}}$. Therefore, the distribution function of heavy ions is more narrow than the proton distribution function but shifted by the same amount due to the wave motion as can be seen in Eq.~(\ref{fwave}). This can lead to a more severe deformation of the distribution function after the averaging process, which might also lead to ring-like apparent VDFs for heavy ions in strong wave fields.

The apparent temperature anisotropy can now be calculated by taking the
second moment of the distribution function according to the formula:
\begin{equation}\label{appaniso}
A=\frac{T_{\perp}}{T_{\parallel}}=\frac{\int\limits_{-\infty}^{+\infty}\int
\limits_{0}^{+\infty}\bar f_1 v_{\perp}^3\mathrm dv_{\perp}\mathrm
dv_{\parallel}}{2\int\limits_{-\infty}^{+\infty}\int
\limits_{0}^{+\infty}\bar f_1v_{\parallel}^2v_{\perp}\mathrm
dv_{\perp}\mathrm dv_{\parallel}}.
\end{equation}
The integration over $\varphi$ leads to the factor 2 in the denominator. The
above distribution function of Figure~\ref{fig_final_linear_deformed_sw} has
an apparent temperature anisotropy of $A=1.64$.

\begin{figure}[t]
\vspace*{2mm}
\begin{center}
\includegraphics[width=8.3cm]{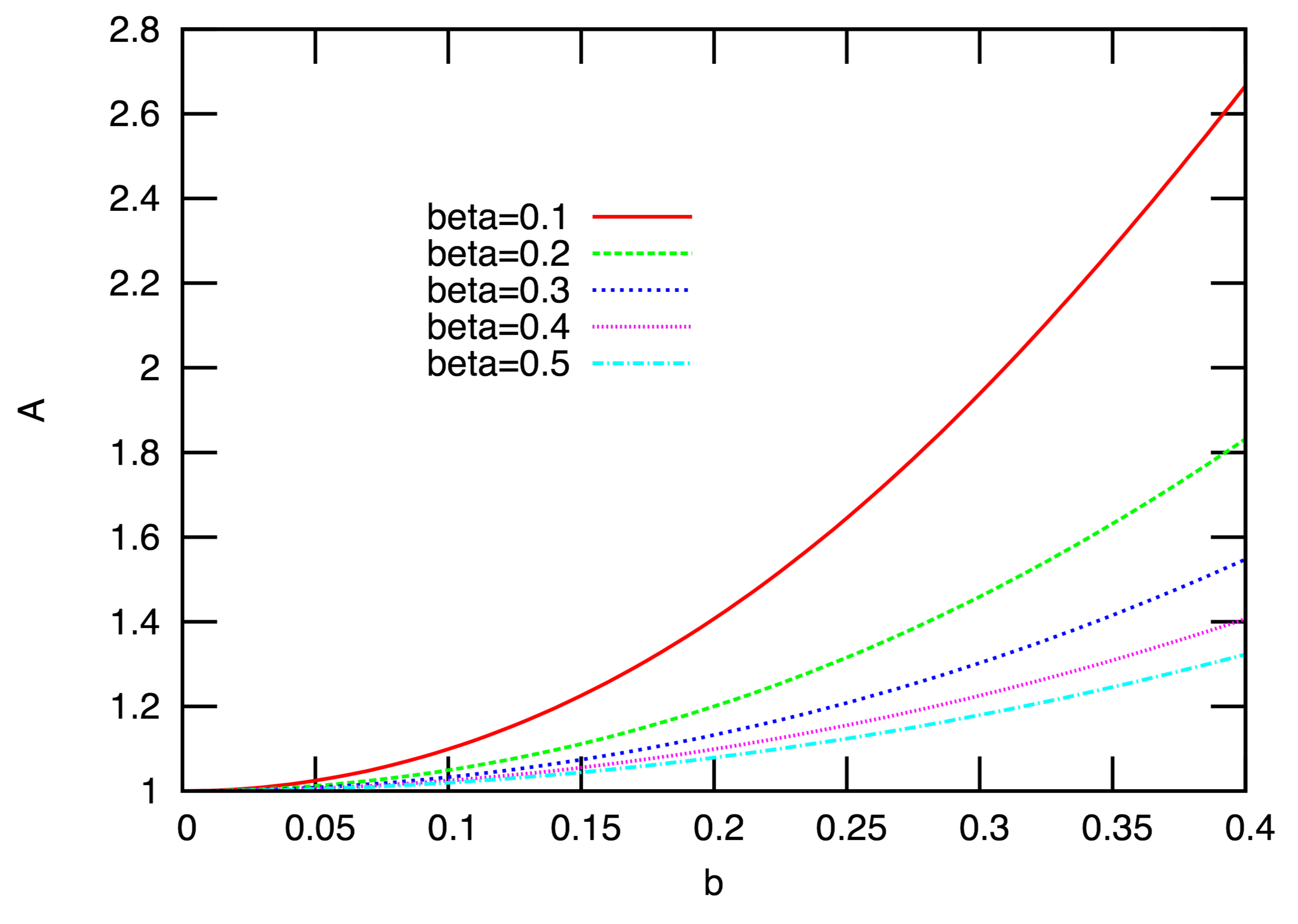}
\end{center}
\caption{Apparent temperature anisotropy in dependence on the normalized
amplitude $b$ of a wave with $\omega=2\pi /T$. The evaluation of the
anisotropy according to Eq.~(\ref{appaniso}) is shown for different values of
the plasma beta. } \label{fig_final_linear_deformed_anisotropy}
\end{figure}

By varying the wave parameters and the particle thermal speed, corresondingly
varying forms of the distribution function can be created. The dependence of
the resulting apparent temperature anisotropy on the plasma beta and the wave
amplitude $b$ is shown in Fig.~\ref{fig_final_linear_deformed_anisotropy}.

\begin{figure}[t]
\vspace*{2mm}
\begin{center}
\includegraphics[width=8.3cm]{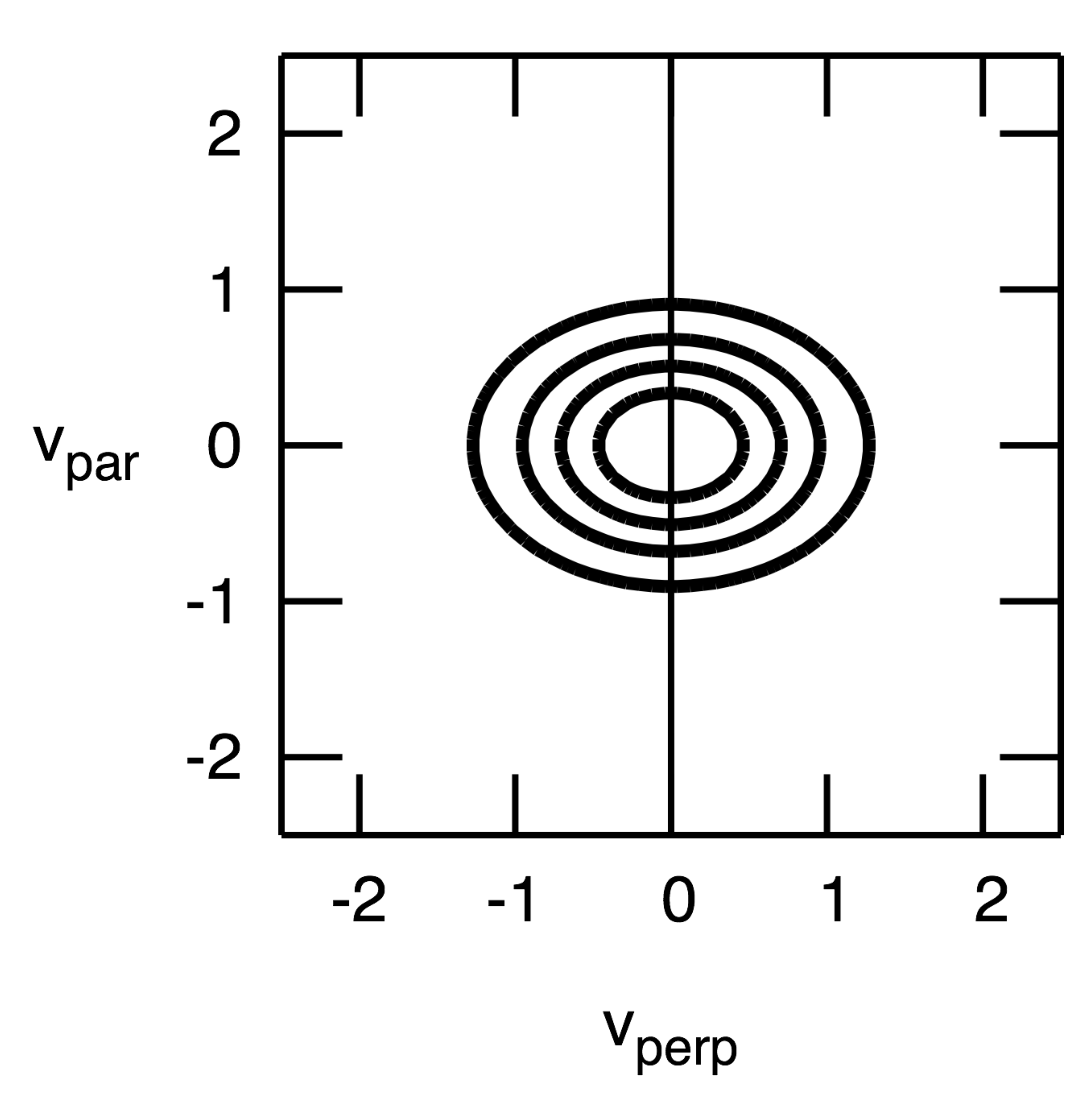}
\end{center}
\caption{Bi-Maxwellian velocity distribution function for a temperature anisotropy of 2.} \label{fig_final_linear_deformed_bimaxwell}
\end{figure}

The obtained distribution function looks from the first point of view very similar to a classical bi-Maxwellian distribution function of the form
\begin{equation}\label{fbm}
f_{\mathrm{bm}}\sim \exp\left(-\frac{v_{\perp}^2}{v_{\mathrm{th}\perp}^2}-\frac{v_{\parallel}^2}{v_{\mathrm{th}\parallel}^2}\right).
\end{equation}
In Fig.~\ref{fig_final_linear_deformed_bimaxwell}, we show a bi-Maxwellian VDF for an (intrinsic) anisotropy of $v_{\mathrm{th}\perp}^2/v_{\mathrm{th}\parallel}^2=2$. 
The general form of the VDF in Fig.~\ref{fig_final_linear_deformed_sw} can also be approximated by this mathematical representation, which underlines the difficulties arising from the correct definition of the observed temperature. Interpreting the wave-broadened distribution function $\bar f$ as a bi-Maxwellian leads to an apparent difference in the two thermal velocities of the bi-Maxwellian VDF. Therefore, the wave-broadening can also be expressed in terms of a corresponding apparent thermal velocity anisotropy. Applying the second moment relation from Eq.~(\ref{appaniso}) to a bi-Maxwellian distribution function yields the ratio $A=v_{\mathrm{th}\perp}^2/v_{\mathrm{th}\parallel}^2$. This means that the anisotropy in the apparent thermal speeds is given directly by $\sqrt A$. 

\begin{figure*}[t]
\vspace*{2mm}
\begin{center}
\includegraphics[width=12cm]{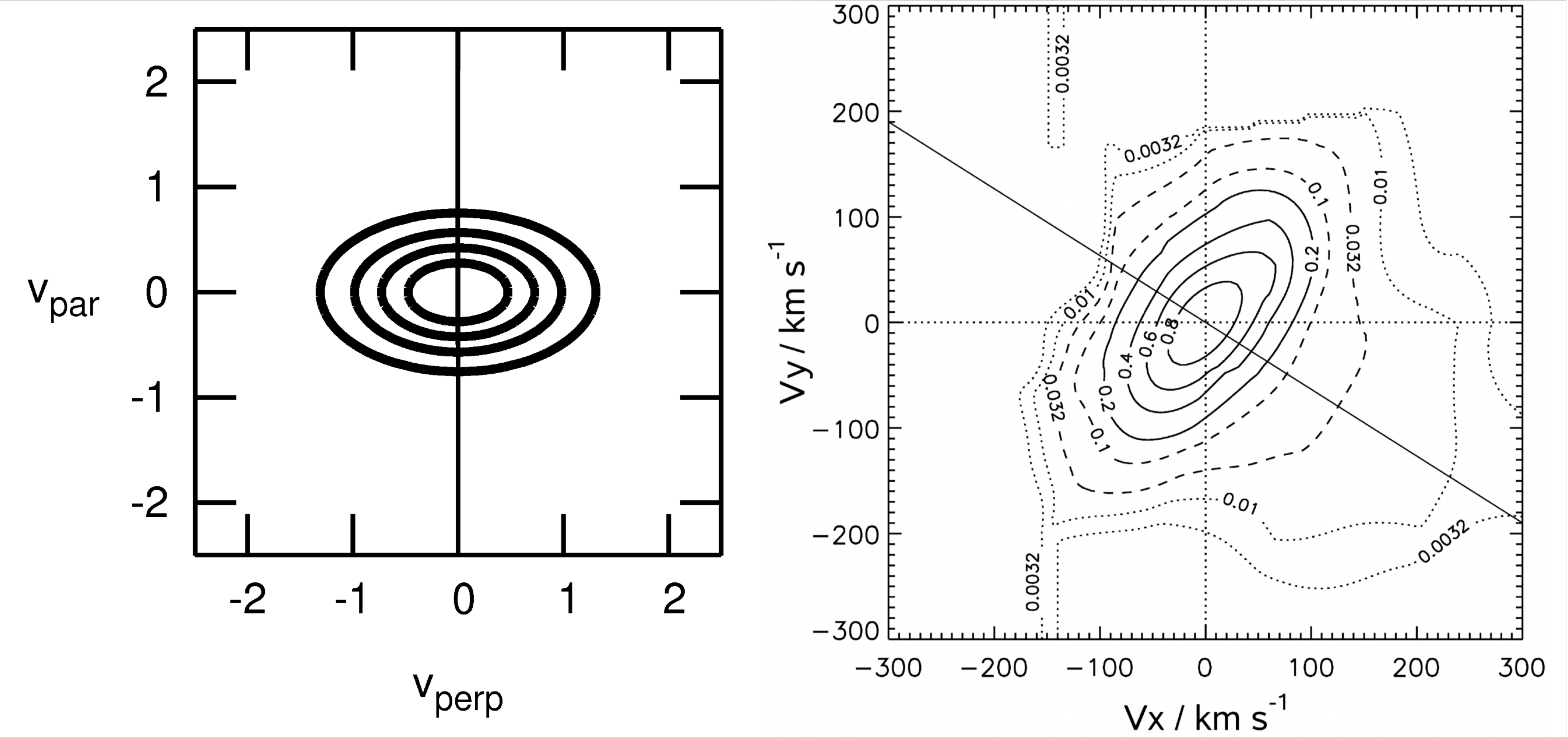}
\end{center}
\caption{\emph{left}: Like in Fig.~\ref{fig_final_linear_deformed_sw} but including higher-order corrections. \emph{right}: Typical proton distribution function measured by the Helios 2 spacecraft in 1976. The black line indicates the direction of the background magnetic field.}
\label{fig_final_linear_deformed_helios}
\end{figure*}

Considering waves with higher frequencies, we must admit that the
approximation $kv_{\parallel}\ll \Omega$ is not valid anymore. Then we have
to accept further modifications of the VDF in the wave field, and we find a
non-Maxwellian dependence also on the parallel speed coordinate
$v_{\parallel}$. Accordingly, the distribution is
\begin{equation}\label{fh}
f_{\mathrm h}=f_{\mathrm w}\exp\left(\frac{2v_{\parallel}}{v_{\mathrm{th}}^2}
\left[\frac{\omega}{k}+\frac{B_0}{b}V_{\mathrm
w}\left(1-\frac{\omega}{\Omega}+\frac{kv_{\parallel}}{2\Omega}\right)\right]\right).
\end{equation}
If we apply the time-averaging process to this distribution function, the
results change significantly. The additional term makes the distribution
function more prolate, already at lower wave amplitude and for higher beta
values. Thus, it becomes even more similar to a bi-Maxwellian distribution function as it is shown in Fig.~\ref{fig_final_linear_deformed_bimaxwell}. An example for this situation is shown in
Fig.~\ref{fig_final_linear_deformed_helios}, where we assumed the parameters
$\omega=10\pi/T$, $b=0.1$, and $\beta=0.5$.

For comparison, also a typical measurement by the Helios 2 spacecraft from 1976 is
shown. The anisotropy of the observed distribution function is comparable to
the calculated apparent anisotropy, whereas other effects such as the formed
beam along the background field is not reproduced by the above calculations.
We find that the model distribution function fits the observed distribution
function well. This shows that the broadening effect and the detailed shaping
mainly depends on the frequency of the waves and the plasma beta. The
measured distribution function is better represented by the corrected
distribution function for higher frequencies. This distribution is not simply
a shifted Maxwellian but has a further non-Maxwellian dependence on
$v_{\parallel}$, which can represent the observations better.

\conclusions

In this study, we have shown how the VDF is shaped by the presence of a
large-amplitude wave. In the case of transversal wave activity, the
distribution function obtains a shift in the direction perpendicular to the background magnetic field. If the
distribution function is averaged over time, this shift will lead to a
smearing in the perpendicular velocity component, which in turn would be
interpreted as a temperature anisotropy in favour of the perpendicular
direction. Every real measured distribution function can only be determined
by sampling within a certain time period, and this implies averaging. Thus
the resulting temperature as the second moment of the VDF reflects this
procedure.

We could demonstrate, using a simplified model, how this effect can lead to a
significant change in the observed distribution functions, as plasma
measurements are always done by counting particles over a certain sampling
time $T$. This sampling period corresponds again to time averaging. The
broadening of particle distributions due to microturbulence is a well-known
fact, which is exploited in spectroscopy to determine remotely, for example,
the turbulence level in the solar corona \citep[e.g.][]{kohl06} from
ultraviolet emission line broadenings. In the context of measurements of
plasma VDF in the solar wind, however, this was not taken into account before
to the authors' knowledge.

Also compressive fluctuations can be treated in a similar way. However, then
broadening would be observed mostly in the parallel direction. Consistently
with the present emphasis on perpendicular broadening, most recent
observations show a higher transversal wave activity in almost all cases in
the fast solar wind \citep{horbury05,alexandrova08b}.

The meaning of intrinsic temperature anisotropies should be further discussed
in the future. As mentioned before, a severe limitation to
$T_{\perp}/T{\parallel}$ is observed in the solar wind 
\citep{marsch06b,bale09, marsch09} in relation with plasma micro-instabilities which reveal a sensitive beta-dependence. This
finding gives a clear indication that the observed temperature anisotropies
are largely intrinsic. This finding even more underlines the importance of an
adequate definition and treatment of the measured plasma temperatures. The
apparent higher temperature that we find in our model VDFs is not the result
of resonant heating processes, such as Landau damping or cyclotron resonant
wave--particle interactions. Hence, this wave-related mechanism is reversible,
and for a vanishing wave field also the apparent anisotropy would disappear.

Nonresonant wave--particle interactions have been studied in the framework of quasi-linear theory by \citet{bourouaine08}. They demonstrate that the nonresonant heating is more effective for lower plasma betas since the efficiency of the quasi-linear diffusion is mainly proportional to $v_{\mathrm A}-v_{\parallel}$.  We can confirm the beta dependence and find the dependence of the parallel particle velocity in the wave frame also in our considerations as in Eq.~(\ref{Pconstant}) for example. However, we could show that the dispersion of Alfv\'en/ion-cyclotron waves can compensate for the reference frame shift in the Vlasov picture. Furthermore, \citet{bourouaine08} found a stronger effect of the nonresonant heating on heavy ion species which is also consistent with our model.

In the context of our measurement effect according to Eq.~(\ref{fbarbegin}), the determination of temperature is not an ergodic measurement anymore if wave fields lead to a coherent particle motion as it is the case in the solar wind. This leads to the problem that the assumption of Markovian statistics is violated since the particle motion is additionally affected by the deterministically time-dependent wave motion during the averaging. Non-ergodicity in our case means that the temperature based on time averaging is different from the temperature based on ensemble averaging. But all real temperature measurements in dilute plasmas have to be done by averaging over time leading to the apparent deformations in the distribution function that we showed above. An appropriate ensemble average, however, is not accesible on the required scales in the solar wind due to the low particle number density.  Recently, \citet{hizanidis10} found that the applicability of classical quasi-linear diffusion is not guaranteed in a coherent electromagnetic wave field. They developed a new kinetic theory for wave--particle interactions and find time-dependent diffusion tensors describing the velocity evolution of the VDF. This is a manifestation of the non-ergodicity of the measurement process. Maybe a completely different description of the microphysical behavior should be applied to these coherent cases as it is proposed and discussed in the textbook by \citet{elskens03}. The wave broadening effect could be excluded locally if, at the position and time where the measurement is taken, enough particles are present with the local coherent speed additional to their thermal speed. In all accessible solar wind plasma cases however, the number of particles that can be counted under constant conditions compared to $\phi=kz-\omega t$ is practically always too small.

A shorter sampling time would bring the observation closer to a ``snapshot''
of the real distribution function. A faster measurement would help avoiding
heavy smearing in velocity space due to wave activity. The comparatively
small number density of particles in the solar wind, however, requires long
sampling times or larger geometry factors which are not affordable. The
Solar Orbiter mission should provide new insights, because the improved
instruments to be flown on board this spacecraft and the higher particle
densities expected closer to the sun will permit much shorter sampling times
down to 100~ms. 

In this work, we assumed a monochromatic wave that leads to smearing out of
the VDF by the time-averaging process. A more realistic assumption would be a
broad spectrum of waves, which the particles have to follow and the VDF to
respond to. Furthermore, the particle flux will not be steady in all
directions during the sampling time $T$. Also, a realistic spacecraft model
should be applied, including the detailed time dependence of the sampling
method as well as the relative velocity of the spacecraft with respect to the
solar wind. If such appropriate models were available, the analysis of the
distribution function could in turn provide new information about the waves,
such as their polarization or propagation direction. Up to now, in our model
only waves propagating parallel to the background magnetic field have been
dealt with. To complete our analysis also oblique wave propagation should be
taken into account.

\paragraph{Acknowledgements}
D.~V. is grateful for financial support by the International Max Planck Research
School (IMPRS) on Physical Processes in the Solar System and Beyond.

\bibliographystyle{copernicus}
\bibliography{paper_linear_deformed_pub}

\end{document}